%% file: main.tex
\documentclass[10pt,conference]{IEEEtran}
\IEEEoverridecommandlockouts

\usepackage{hyperref}
\usepackage{cite}
\usepackage[table,xcdraw]{xcolor} % For coloring cells in the table
\usepackage{amsmath,amssymb,amsfonts}
\usepackage{textcomp}
\usepackage{amsmath}
\usepackage{comment}
\usepackage{tabularray}
\usepackage[longtable]{multirow}
\usepackage{colortbl}
\usepackage{longtable}
\usepackage{tcolorbox}
\usepackage{hhline}
\usepackage{comment}
\usepackage{algorithm}
\usepackage{algpseudocode}
\usepackage{algorithmicx}
\usepackage[noend]{algcompatible}
\usepackage{multirow} % For multirow cells in the table
\usepackage{graphicx} % For \resizebox to adjust table width
\usepackage{listings}
\usepackage{tikz}
\usetikzlibrary{calc}
\usetikzlibrary{shapes.geometric}
\usetikzlibrary{decorations.pathreplacing}
\usetikzlibrary{positioning}
\usetikzlibrary{shapes}
\usepackage[normalem]{ulem} % normalem do not replace \emph to underline
\usepackage{xspace}
\usepackage{booktabs}
\usepackage{enumitem}
\setlist[itemize]{leftmargin=*}
\setlist[enumerate]{leftmargin=*}

\input{defs/defs-code}

\usepackage{array}
\usepackage{ragged2e}

\usepackage{caption}

\makeatletter
\newcommand{\DefMacro}{\@ifstar\@DefMacroAllowRedefine\@DefMacro}
\newcommand{\@DefMacro}[2]{\expandafter\newcommand\csname rmk-#1\endcsname{#2}}
\newcommand{\@DefMacroAllowRedefine}[2]{\expandafter\providecommand\csname rmk-#1\endcsname{} \expandafter\renewcommand\csname rmk-#1\endcsname{#2}}
\makeatother
\newcommand{\UseMacro}[1]{\csname rmk-#1\endcsname}

\makeatletter
    \newcommand{\coloruwave}[2]{%
        \UL@protected\def\temp@uwave{\leavevmode \bgroup 
        \ifdim \ULdepth=\maxdimen \ULdepth 3.5\p@
        \else \advance\ULdepth2\p@ 
        \fi \markoverwith{\textcolor{#1}{\lower\ULdepth\hbox{\sixly \char58}}}\ULon}
        \font\sixly=lasy6 % does not re-load if already loaded, so no memory drain.
        \temp@uwave{#2}%
    }
\makeatother

\newcommand*\circled[1]{\kern-2.5em%
  \put(0,4){\color{black}\circle*{10}}\put(0,4){\circle{8}}%
  \put(-1.8,1.5){\color{white}\bfseries\scriptsize#1}~~}
\usepackage{enumitem}

\def\BibTeX{{\rm B\kern-.05em{\sc i\kern-.025em b}\kern-.08em
    T\kern-.1667em\lower.7ex\hbox{E}\kern-.125emX}}

\newcommand{\intertrans}{\textsc{InterTrans}\xspace}

\makeatletter
\newcommand{\linebreakand}{%
  \end{@IEEEauthorhalign}
  \hfill\mbox{}\par
  \mbox{}\hfill\begin{@IEEEauthorhalign}
}
\makeatother

\begin{document}

\title{\fontsize{21}{24}\selectfont \intertrans: Leveraging Transitive Intermediate Translations to Enhance LLM-based Code Translation}

\author{\IEEEauthorblockN{1\textsuperscript{st} Marcos Macedo}
\IEEEauthorblockA{\textit{School of Computing} \\
\textit{Queen's University}\\
Kingston, ON, Canada \\
marcos.macedo@queensu.ca}
\and
\IEEEauthorblockN{2\textsuperscript{nd} Yuan Tian}
\IEEEauthorblockA{\textit{School of Computing} \\
\textit{Queen's University}\\
Kingston, ON, Canada \\
y.tian@queensu.ca}
\and
\IEEEauthorblockN{3\textsuperscript{rd} Pengyu Nie}
\IEEEauthorblockA{\textit{Cheriton School of Computer Science} \\
\textit{University of Waterloo}\\
Waterloo, ON, Canada \\
pynie@uwaterloo.ca}
\linebreakand
\IEEEauthorblockN{4\textsuperscript{th} Filipe R. Cogo}
\IEEEauthorblockA{\textit{Centre for Software Excellence} \\
\textit{Huawei Canada}\\
Kingston, ON, Canada \\
filipe.roseiro.cogo1@huawei.com}
\and
\IEEEauthorblockN{5\textsuperscript{th} Bram Adams}
\IEEEauthorblockA{\textit{School of Computing} \\
\textit{Queen's University}\\
Kingston, ON, Canada \\
bram.adams@queensu.ca}
}

\maketitle

\begin{abstract}
Code translation aims to convert a program from one programming language (PL) to another. This long-standing software engineering task is crucial for modernizing legacy systems, ensuring cross-platform compatibility, enhancing performance, and more. However, automating this process remains challenging due to many syntactic and semantic differences between PLs. Recent studies show that even advanced techniques such as large language models (LLMs), especially open-source LLMs, still struggle with the task.

Currently, code LLMs are trained with source code from multiple programming languages, thus presenting multilingual capabilities. In this paper, we investigate whether such capabilities can be harnessed to enhance code translation. To achieve this goal, we introduce \intertrans, an LLM-based automated code translation approach that, in contrast to existing approaches, leverages intermediate translations to bridge the syntactic and semantic gaps between source and target PLs. \intertrans contains two stages. It first utilizes a novel Tree of Code Translation (ToCT) algorithm to plan transitive intermediate translation sequences between a given source and target PL, then validates them in a specific order. We evaluate \intertrans with three open LLMs on three benchmarks (i.e., CodeNet, HumanEval-X, and TransCoder) involving six PLs. Results show an \textit{absolute improvement} of 18.3\% to 43.3\% in Computation Accuracy (CA) for \intertrans over Direct Translation with 10 attempts. The best-performing variant of \intertrans (with the Magicoder LLM) achieved an average CA of 87.3\%-95.4\% on three benchmarks.

\end{abstract}

\IEEEpeerreviewmaketitle

\section{Introduction}\label{sec:intro}
\input{intro}

\section{InterTrans}\label{sec:intertrans}
\input{method}

\section{Experiment Design}\label{sec:exp}
\input{exp}

\section{Results and Analysis}\label{sec:result}
\input{results}

\section{Discussion}\label{sec:discussion}
\input{discuss}

\section{Related Work}\label{sec:related}
\input{related}

\section{Conclusion}\label{sec:conclude}
\input{conclude}

\section{Acknowledgments} 
We express our gratitude to the Natural Sciences and Engineering Research Council of Canada (NSERC) for their support, with funding reference number RGPIN-2019-05071. Additionally, we extend our appreciation to the Vector Institute for its offering of the Vector Scholarship in Artificial Intelligence, which was awarded to the first author. The findings and opinions expressed in this paper are those of the authors and do not necessarily represent or reflect those of Huawei and/or its subsidiaries and affiliates.

\clearpage
\bibliography{reference}
\bibliographystyle{plain}

\end{document}

%% file: defs/defs-code.tex
\definecolor{gray}{RGB}{211,211,211}
\newcommand{\jbasicstyle}{\small\sffamily} % Style of code

\newcommand{\jnumberstyle}{\scriptsize}

\lstdefinelanguage{pseudo}
{
  morekeywords={},
  keywordstyle=\bfseries,
  lineskip=-0.1em,
  numbers=left, % none for no numbers
  numberstyle=\jnumberstyle,
  numbersep=4pt,
  basicstyle=\jbasicstyle,
  breaklines=true,
  breakautoindent=true,
  tabsize=2,
  columns=fullflexible,
  morecomment=*[l][\textsl]{//},
  mathescape=true,
  xleftmargin=10pt,
}

\lstdefinelanguage{todo-comment}
{
  morekeywords={},
  keywordstyle=\bfseries,
  lineskip=-0.1em,
  numbers=none,
  basicstyle=\jbasicstyle,
  breaklines=true,
  breakautoindent=true,
  tabsize=2,
  columns=fullflexible,
  morecomment=*[l][\textsl]{//},
  mathescape=true,
  xleftmargin=-10pt,
}

\lstdefinelanguage{java-pretty}
{
  language=java,
  numbers=left,
  basicstyle=\scriptsize\ttfamily,
  numberstyle=\scriptsize,
  breaklines=true,
  columns=fullflexible,
  xleftmargin=16pt,
  showstringspaces=false,
}

\lstdefinelanguage{java-pretty-no-number}
{
  language=java,
  numbers=none,
  basicstyle=\scriptsize\ttfamily,
  numberstyle=\scriptsize,
  breaklines=true,
  columns=fullflexible,
  xleftmargin=16pt,
  showstringspaces=false,
}

\lstdefinelanguage{Rust}{%
  sensitive%
, morecomment=[l]{//}%
, morecomment=[s]{/*}{*/}%
, moredelim=[s][{\itshape\color[rgb]{0,0,0.75}}]{\#[}{]}%
, morestring=[b]{"}%
, alsodigit={}%
, alsoother={}%
, alsoletter={!}%
, morekeywords={break, continue, else, for, if, in, loop, match, return, while}  % control flow keywords
, morekeywords={as, const, let, move, mut, ref, static}  % in the context of variables
, morekeywords={dyn, enum, fn, impl, Self, self, struct, trait, type, union, use, where}  % in the context of declarations
, morekeywords={crate, extern, mod, pub, super}  % in the context of modularisation
, morekeywords={unsafe}  % markers
, morekeywords={abstract, alignof, become, box, do, final, macro, offsetof, override, priv, proc, pure, sizeof, typeof, unsized, virtual, yield}  % reserved identifiers
, morekeywords=[2]{Add, AddAssign, Any, AsciiExt, AsInner, AsInnerMut, AsMut, AsRawFd, AsRawHandle, AsRawSocket, AsRef, Binary, BitAnd, BitAndAssign, Bitor, BitOr, BitOrAssign, BitXor, BitXorAssign, Borrow, BorrowMut, Boxed, BoxPlace, BufRead, BuildHasher, CastInto, CharExt, Clone, CoerceUnsized, CommandExt, Copy, Debug, DecodableFloat, Default, Deref, DerefMut, DirBuilderExt, DirEntryExt, Display, Div, DivAssign, DoubleEndedIterator, DoubleEndedSearcher, Drop, EnvKey, Eq, Error, ExactSizeIterator, ExitStatusExt, Extend, FileExt, FileTypeExt, Float, Fn, FnBox, FnMut, FnOnce, Freeze, From, FromInner, FromIterator, FromRawFd, FromRawHandle, FromRawSocket, FromStr, FullOps, FusedIterator, Generator, Hash, Hasher, Index, IndexMut, InPlace, Int, Into, IntoCow, IntoInner, IntoIterator, IntoRawFd, IntoRawHandle, IntoRawSocket, IsMinusOne, IsZero, Iterator, JoinHandleExt, LargeInt, LowerExp, LowerHex, MetadataExt, Mul, MulAssign, Neg, Not, Octal, OpenOptionsExt, Ord, OsStrExt, OsStringExt, Packet, PartialEq, PartialOrd, Pattern, PermissionsExt, Place, Placer, Pointer, Product, Put, RangeArgument, RawFloat, Read, Rem, RemAssign, Seek, Shl, ShlAssign, Shr, ShrAssign, Sized, SliceConcatExt, SliceExt, SliceIndex, Stats, Step, StrExt, Sub, SubAssign, Sum, Sync, TDynBenchFn, Terminal, Termination, ToOwned, ToSocketAddrs, ToString, Try, TryFrom, TryInto, UnicodeStr, Unsize, UpperExp, UpperHex, WideInt, Write}
, morekeywords=[2]{Send}  % additional traits
, morekeywords=[3]{bool, char, f32, f64, i8, i16, i32, i64, isize, str, u8, u16, u32, u64, unit, usize, i128, u128}  % primitive types
, morekeywords=[4]{Err, false, None, Ok, Some, true}  % prelude value constructors
, morekeywords=[3]{AccessError, Adddf3, AddI128, AddoI128, AddoU128, ADDRESS, ADDRESS64, addrinfo, ADDRINFOA, AddrParseError, Addsf3, AddU128, advice, aiocb, Alignment, AllocErr, AnonPipe, Answer, Arc, Args, ArgsInnerDebug, ArgsOs, Argument, Arguments, ArgumentV1, Ashldi3, Ashlti3, Ashrdi3, Ashrti3, AssertParamIsClone, AssertParamIsCopy, AssertParamIsEq, AssertUnwindSafe, AtomicBool, AtomicPtr, Attr, auxtype, auxv, BackPlace, BacktraceContext, Barrier, BarrierWaitResult, Bencher, BenchMode, BenchSamples, BinaryHeap, BinaryHeapPlace, blkcnt, blkcnt64, blksize, BOOL, boolean, BOOLEAN, BoolTrie, BorrowError, BorrowMutError, Bound, Box, bpf, BTreeMap, BTreeSet, Bucket, BucketState, Buf, BufReader, BufWriter, Builder, BuildHasherDefault, BY, BYTE, Bytes, CannotReallocInPlace, cc, Cell, Chain, CHAR, CharIndices, CharPredicateSearcher, Chars, CharSearcher, CharsError, CharSliceSearcher, CharTryFromError, Child, ChildPipes, ChildStderr, ChildStdin, ChildStdio, ChildStdout, Chunks, ChunksMut, ciovec, clock, clockid, Cloned, cmsgcred, cmsghdr, CodePoint, Color, ColorConfig, Command, CommandEnv, Component, Components, CONDITION, condvar, Condvar, CONSOLE, CONTEXT, Count, Cow, cpu, CRITICAL, CStr, CString, CStringArray, Cursor, Cycle, CycleIter, daddr, DebugList, DebugMap, DebugSet, DebugStruct, DebugTuple, Decimal, Decoded, DecodeUtf16, DecodeUtf16Error, DecodeUtf8, DefaultEnvKey, DefaultHasher, dev, device, Difference, Digit32, DIR, DirBuilder, dircookie, dirent, dirent64, DirEntry, Discriminant, DISPATCHER, Display, Divdf3, Divdi3, Divmoddi4, Divmodsi4, Divsf3, Divsi3, Divti3, dl, Dl, Dlmalloc, Dns, DnsAnswer, DnsQuery, dqblk, Drain, DrainFilter, Dtor, Duration, DwarfReader, DWORD, DWORDLONG, DynamicLibrary, Edge, EHAction, EHContext, Elf32, Elf64, Empty, EmptyBucket, EncodeUtf16, EncodeWide, Entry, EntryPlace, Enumerate, Env, epoll, errno, Error, ErrorKind, EscapeDebug, EscapeDefault, EscapeUnicode, event, Event, eventrwflags, eventtype, ExactChunks, ExactChunksMut, EXCEPTION, Excess, ExchangeHeapSingleton, exit, exitcode, ExitStatus, Failure, fd, fdflags, fdsflags, fdstat, ff, fflags, File, FILE, FileAttr, filedelta, FileDesc, FilePermissions, filesize, filestat, FILETIME, filetype, FileType, Filter, FilterMap, Fixdfdi, Fixdfsi, Fixdfti, Fixsfdi, Fixsfsi, Fixsfti, Fixunsdfdi, Fixunsdfsi, Fixunsdfti, Fixunssfdi, Fixunssfsi, Fixunssfti, Flag, FlatMap, Floatdidf, FLOATING, Floatsidf, Floatsisf, Floattidf, Floattisf, Floatundidf, Floatunsidf, Floatunsisf, Floatuntidf, Floatuntisf, flock, ForceResult, FormatSpec, Formatted, Formatter, Fp, FpCategory, fpos, fpos64, fpreg, fpregset, FPUControlWord, Frame, FromBytesWithNulError, FromUtf16Error, FromUtf8Error, FrontPlace, fsblkcnt, fsfilcnt, fsflags, fsid, fstore, fsword, FullBucket, FullBucketMut, FullDecoded, Fuse, GapThenFull, GeneratorState, gid, glob, glob64, GlobalDlmalloc, greg, group, GROUP, Guard, GUID, Handle, HANDLE, Handler, HashMap, HashSet, Heap, HINSTANCE, HMODULE, hostent, HRESULT, id, idtype, if, ifaddrs, IMAGEHLP, Immut, in, in6, Incoming, Infallible, Initializer, ino, ino64, inode, input, InsertResult, Inspect, Instant, int16, int32, int64, int8, integer, IntermediateBox, Internal, Intersection, intmax, IntoInnerError, IntoIter, IntoStringError, intptr, InvalidSequence, iovec, ip, IpAddr, ipc, Ipv4Addr, ipv6, Ipv6Addr, Ipv6MulticastScope, Iter, IterMut, itimerspec, itimerval, jail, JoinHandle, JoinPathsError, KDHELP64, kevent, kevent64, key, Key, Keys, KV, l4, LARGE, lastlog, launchpad, Layout, Lazy, lconv, Leaf, LeafOrInternal, Lines, LinesAny, LineWriter, linger, linkcount, LinkedList, load, locale, LocalKey, LocalKeyState, Location, lock, LockResult, loff, LONG, lookup, lookupflags, LookupHost, LPBOOL, LPBY, LPBYTE, LPCSTR, LPCVOID, LPCWSTR, LPDWORD, LPFILETIME, LPHANDLE, LPOVERLAPPED, LPPROCESS, LPPROGRESS, LPSECURITY, LPSTARTUPINFO, LPSTR, LPVOID, LPWCH, LPWIN32, LPWSADATA, LPWSAPROTOCOL, LPWSTR, Lshrdi3, Lshrti3, lwpid, M128A, mach, major, Map, mcontext, Metadata, Metric, MetricMap, mflags, minor, mmsghdr, Moddi3, mode, Modsi3, Modti3, MonitorMsg, MOUNT, mprot, mq, mqd, msflags, msghdr, msginfo, msglen, msgqnum, msqid, Muldf3, Mulodi4, Mulosi4, Muloti4, Mulsf3, Multi3, Mut, Mutex, MutexGuard, MyCollection, n16, NamePadding, NativeLibBoilerplate, nfds, nl, nlink, NodeRef, NoneError, NonNull, NonZero, nthreads, NulError, OccupiedEntry, off, off64, oflags, Once, OnceState, OpenOptions, Option, Options, OptRes, Ordering, OsStr, OsString, Output, OVERLAPPED, Owned, Packet, PanicInfo, Param, ParseBoolError, ParseCharError, ParseError, ParseFloatError, ParseIntError, ParseResult, Part, passwd, Path, PathBuf, PCONDITION, PCONSOLE, Peekable, PeekMut, Permissions, PhantomData, pid, Pipes, PlaceBack, PlaceFront, PLARGE, PoisonError, pollfd, PopResult, port, Position, Powidf2, Powisf2, Prefix, PrefixComponent, PrintFormat, proc, Process, PROCESS, processentry, protoent, PSRWLOCK, pthread, ptr, ptrdiff, PVECTORED, Queue, radvisory, RandomState, Range, RangeFrom, RangeFull, RangeInclusive, RangeMut, RangeTo, RangeToInclusive, RawBucket, RawFd, RawHandle, RawPthread, RawSocket, RawTable, RawVec, Rc, ReadDir, Receiver, recv, RecvError, RecvTimeoutError, ReentrantMutex, ReentrantMutexGuard, Ref, RefCell, RefMut, REPARSE, Repeat, Result, Rev, Reverse, riflags, rights, rlim, rlim64, rlimit, rlimit64, roflags, Root, RSplit, RSplitMut, RSplitN, RSplitNMut, RUNTIME, rusage, RwLock, RWLock, RwLockReadGuard, RwLockWriteGuard, sa, SafeHash, Scan, sched, scope, sdflags, SearchResult, SearchStep, SECURITY, SeekFrom, segment, Select, SelectionResult, sem, sembuf, send, Sender, SendError, servent, sf, Shared, shmatt, shmid, ShortReader, ShouldPanic, Shutdown, siflags, sigaction, SigAction, sigevent, sighandler, siginfo, Sign, signal, signalfd, SignalToken, sigset, sigval, Sink, SipHasher, SipHasher13, SipHasher24, size, SIZE, Skip, SkipWhile, Slice, SmallBoolTrie, sockaddr, SOCKADDR, sockcred, Socket, SOCKET, SocketAddr, SocketAddrV4, SocketAddrV6, socklen, speed, Splice, Split, SplitMut, SplitN, SplitNMut, SplitPaths, SplitWhitespace, spwd, SRWLOCK, ssize, stack, STACKFRAME64, StartResult, STARTUPINFO, stat, Stat, stat64, statfs, statfs64, StaticKey, statvfs, StatVfs, statvfs64, Stderr, StderrLock, StderrTerminal, Stdin, StdinLock, Stdio, StdioPipes, Stdout, StdoutLock, StdoutTerminal, StepBy, String, StripPrefixError, StrSearcher, subclockflags, Subdf3, SubI128, SuboI128, SuboU128, subrwflags, subscription, Subsf3, SubU128, Summary, suseconds, SYMBOL, SYMBOLIC, SymmetricDifference, SyncSender, sysinfo, System, SystemTime, SystemTimeError, Take, TakeWhile, tcb, tcflag, TcpListener, TcpStream, TempDir, TermInfo, TerminfoTerminal, termios, termios2, TestDesc, TestDescAndFn, TestEvent, TestFn, TestName, TestOpts, TestResult, Thread, threadattr, threadentry, ThreadId, tid, time, time64, timespec, TimeSpec, timestamp, timeval, timeval32, timezone, tm, tms, ToLowercase, ToUppercase, TraitObject, TryFromIntError, TryFromSliceError, TryIter, TryLockError, TryLockResult, TryRecvError, TrySendError, TypeId, U64x2, ucontext, ucred, Udivdi3, Udivmoddi4, Udivmodsi4, Udivmodti4, Udivsi3, Udivti3, UdpSocket, uid, UINT, uint16, uint32, uint64, uint8, uintmax, uintptr, ulflags, ULONG, ULONGLONG, Umoddi3, Umodsi3, Umodti3, UnicodeVersion, Union, Unique, UnixDatagram, UnixListener, UnixStream, Unpacked, UnsafeCell, UNWIND, UpgradeResult, useconds, user, userdata, USHORT, Utf16Encoder, Utf8Error, Utf8Lossy, Utf8LossyChunk, Utf8LossyChunksIter, utimbuf, utmp, utmpx, utsname, uuid, VacantEntry, Values, ValuesMut, VarError, Variables, Vars, VarsOs, Vec, VecDeque, vm, Void, WaitTimeoutResult, WaitToken, wchar, WCHAR, Weak, whence, WIN32, WinConsole, Windows, WindowsEnvKey, winsize, WORD, Wrapping, wrlen, WSADATA, WSAPROTOCOL, WSAPROTOCOLCHAIN, Wtf8, Wtf8Buf, Wtf8CodePoints, xsw, xucred, Zip, zx}
, morekeywords=[5]{assert!, assert_eq!, assert_ne!, cfg!, column!, compile_error!, concat!, concat_idents!, debug_assert!, debug_assert_eq!, debug_assert_ne!, env!, eprint!, eprintln!, file!, format!, format_args!, include!, include_bytes!, include_str!, line!, module_path!, option_env!, panic!, print!, println!, select!, stringify!, thread_local!, try!, unimplemented!, unreachable!, vec!, write!, writeln!}  % prelude macros
}%

%% file: intro.tex
Automatically translating source code between different programming languages (PLs) can significantly reduce the time and effort required for software development teams. In the literature, researchers have proposed various automated code translation methods. Data-driven learning-based approaches~\cite{roziere2021leveraging,szafraniec_code_2023} have shown impressive improvements over traditional rule-based methods~\cite{c2rust,cxgo,sharpen}. Unlike rule-based approaches, which rely on handcrafted rules and program analysis techniques, learning-based methods can automatically learn syntactic and semantic patterns from large-scale code repositories. 

Large language models (LLMs) represent the most advanced learning-based approaches developed in recent years and have demonstrated promising results across various software engineering tasks~\cite{fan2023large}. Pre-trained on vast amounts of code (across dozens of PLs) and text data, and equipped with billions of parameters, LLMs can be applied directly to code translation without the need for task-specific continuous training/fine-tuning. This would eliminate the need for costly and time-consuming processes involved in collecting training datasets and developing specialized models for code translation. 

However, recent studies have shown that the performance of LLM-based automated code translation, particularly with open-source LLMs, is still far from the production level, with correct translations ranging from 2.1\% to 47.3\%~\cite{pan_understanding_2023, yang2024exploring}. These studies found that many errors in LLM-generated code translations stem from the models' lack of understanding of syntactic and semantic discrepancies between source and target languages, which can vary significantly across different pairs. For instance, 80\% of the errors in translating from C++ to Go are due to syntactic and semantic differences, while only 23.1\% of such errors occur when translating from C++ to C~\cite{pan_understanding_2023}. This variation is intuitive, as certain PLs naturally share more similarities in syntax and semantics than others. 

A similar phenomenon has been observed in machine translation for human languages, where translating between certain languages is easier than others~\cite{kolovratnik2009statistical}. To improve translations for challenging language pairs, a common strategy is to use parallel corpora with a pivot (bridge) language~\cite{kim2019pivot}. In fact, traditional statistical machine translation between non-English languages, such as French to German, often involves pivoting through English~\cite{wu2007pivot}. This approach remains effective with the rise of multilingual neural machine translation models. For instance, in a recent work by Meta~\cite{fan2021beyond}, training language pairs were collected based on linguistic families and bridge languages, facilitating translation across numerous language pairs without exhaustively mining every possible pair.

Inspired by this idea, this paper explores the potential of leveraging transitive intermediate translations from a source PL into other PLs before translating to the desired target PL, an idea not previously explored in the field of automated code translation. For example, to translate a program written in Python to Java, we might first translate it from Python to C++ and then from C++ to Java, as illustrated in Figure~\ref{fig:tree-example}. This process is done through prompting, without additional training data, thanks to code LLMs that are pre-trained on text and code across multiple PLs and naturally possess multilingual capabilities. While this idea is inspired by machine translation, its potential in the inference stage of LLM-based translation approaches has not been explored. Despite the conceptual simplicity of the idea, a major challenge to address is the choice of the number and type of intermediate language(s), since the optimal choice might be different for each pair of PLs or even each pair of code snippets.

The idea of utilizing existing PLs as ``bridges'' is different than earlier work, TransCoder-IR~\cite{szafraniec_code_2023}, a non-LLM learning-based method that enhances source code pairs by incorporating their corresponding low-level, language-agnostic compiler Intermediate Representations (IR), such as LLVM IRs~\cite{lattner2004llvm}, into the training dataset. Instead of relying on one unified IR to bridge any pair of cross-PL translations, we systematically explore different potential transitive intermediate translations using multiple existing PLs.

\intertrans, our novel LLM-based code translation approach that enhances source-target translations via transitive intermediate translations, operates in two stages. In the first stage, a method called Tree of Code Translations (ToCT) generates a \textit{translation tree} containing all potential translation paths for a specific source-target PL pair, conditioned to a set of pre-defined intermediate PLs and the maximum number of intermediate translations to be explored. In the second stage, translation paths are turned into LLM prompts that are executed in a breadth-first order. \intertrans then uses a readily available test suite to validate whether the generated translation to the target language is correct, enabling early termination of translation path exploration if a successful path is found before completely exploring the translation tree.

To evaluate the effectiveness of \intertrans, we conducted experiments using three code LLMs (Code Llama~\cite{roziere2023code}, Magicoder~\cite{wei_magicoder_2023}, and StarCoder2~\cite{lozhkov2024starcoder}) on 4,926 \textit{translation problems} sourced from three datasets, i.e., CodeNet~\cite{puri2021codenet}, HumanEval-X~\cite{zheng2023codegeex}, and TransCoder~\cite{roziere2020unsupervised}. Each translation problem aims to translate a program writing in a source PL to a target PL. These problems involve 30 different source-target PL pairs across six languages: C++, JavaScript, Java, Python, Go, and Rust. Our results show that \intertrans consistently outperforms direct translation (i.e., without intermediate language translation) with 10 attempts, achieving an absolute Computational Accuracy (CA) improvement of 18.3\% to 43.3\% (median: 28.6\%) across the three LLMs and datasets. Through ablation studies, we analyzed the effects of varying the number and selection of intermediate languages on \intertrans's performance. Generally, increasing the number of intermediate translations enhances CA, though the benefits taper off after three translations. Similarly, incorporating more intermediate languages is advantageous, but gains slow after including three languages. The effectiveness of specific intermediate PLs varies across translation pairs, with notable patterns observed in translations from C++/Python to Java via Rust and from Rust to Go via C++. The main contributions of this paper are as follows:

\begin{itemize}%[label=\protect\circled{\arabic*}]
\item We present the first study demonstrating that intermediate translations based on existing PLs can enhance the performance of LLM-based code translation.
\item We propose ToCT, a novel planning algorithm designed to explore intermediate translations effectively. We also introduce \intertrans, an LLM-based code translation approach that uses ToCT and is orthogonal to existing approaches for code translation.
\item We conducted a comprehensive empirical study to evaluate \intertrans. Our results highlight the effectiveness of \intertrans in enhancing LLM-based code translation. We also provide insights for the practical application of \intertrans.
\end{itemize}

\noindent The code for implementing \intertrans, the datasets, and the notebooks for generating the experiment results are available at: \url{https://github.com/RISElabQueens/InterTrans}.

%% file: method.tex
\intertrans translates programs from a source to a target language using an LLM and a series of transitive intermediate translations. The input of \intertrans includes: (1) a LLM, (2) a program $P_s$ written in a source language $L_s$, (3) the target language $L_t$, (4) a non-empty intermediate PL set $L$ which contains $L_s$ but excludes $L_t$, (5) a hyper-parameter $maxDepth$, which determines the maximum number of transitive intermediate translations. \intertrans utilizes a readily available test suite to evaluate the accuracy of the generated program(s) $TP$ written in the target language, i.e., $TP = \{P_t | P_t \in P_{P_s,L_t} \wedge s \neq t\}$, where $P_{P_s,L_t}$ is the set of programs written in $L_t$ that represent translation candidates for $P_s$.

Given a translation problem aimed at converting a source program $P_s$ into a target language $L_t$, \intertrans operates in two stages. In Stage 1, it constructs all possible \textit{translation (PL) paths} 
using a novel approach called the \textit{Tree of Code Translations (ToCT)}, which identifies potential sequences of transitive translations from $L_s$ to $L_t$ via intermediate languages from the set $L$. Stage 2 then uses the source program $P_s$ and the PL paths generated from Stage 1 to perform inferences with an LLM to generate a set of target programs $TP$ written in $L_t$. These programs, each corresponding to a translation path, are generated and verified sequentially against a test suite. The algorithm terminates when a successful translation is identified, indicated by a $P_t$ that passes the test suite. The following subsections provide detailed descriptions of each stage, accompanied by a running example.

%\BA{font is small} 
\begin{figure*}[h]
    \centering
    \includegraphics[width=\linewidth]{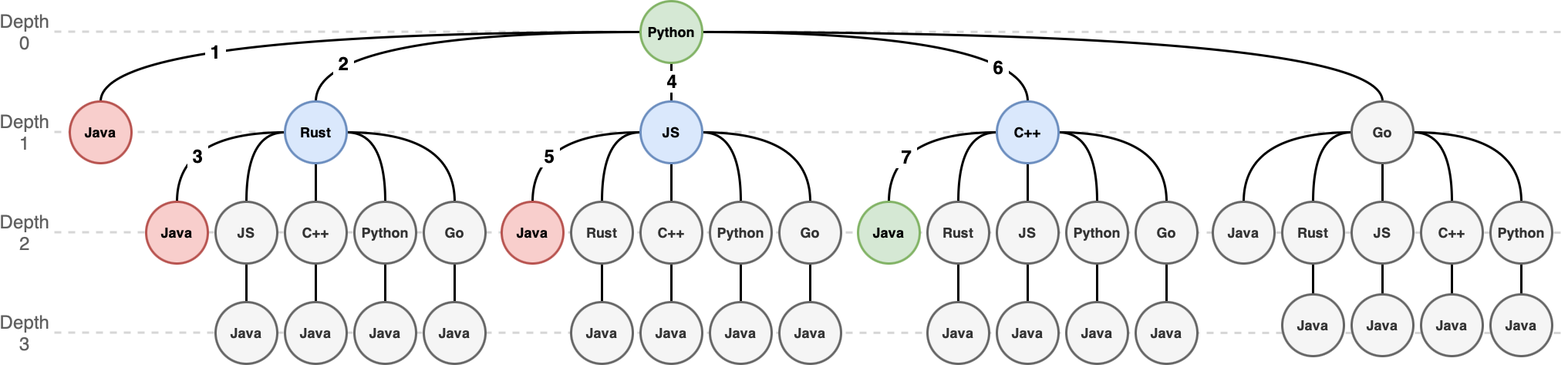} 
\caption{Running example of \intertrans with maxDepth=3 for translating Python to Java, showing a successful translation through C++ after exploring various translation paths. Red nodes represent unsuccessful translations, blue nodes indicate explored translations, green nodes denote successful translations, and grey nodes are skipped translations. The number along with each edge is the execution order of the translations.}
    \label{fig:tree-example}
    \vspace{-10pt} % Reduce space below the figure
\end{figure*}

\subsection{Stage 1: Generating Tree of Code Translations (ToCT)}

Algorithm~\ref{alg:ToCT} specifies how ToCT creates (plans) translation PL paths for a given translation PL pair utilizing a set of intermediate languages. Since ToCT operates at the level of translation PL pairs, this planning algorithm only needs to run once for all translation problems involving the same source and target languages.

In ToCT, the intermediate language set $L$ includes the source language $L_s$ but excludes the target language $L_t$. This is because $L_t$ should be the final target and should not occur as an intermediate step in the translation process, while we should allow $L_s$ to appear in intermediate translations (for cases where a source program can be ``simplified'' by translating to and from another PL). Below, we use a running example, shown in Figure~\ref{fig:tree-example}, to illustrate this algorithm. In this example, we aim to translate a Python program to Java ($L_s$ is Python, $L_t$ is Java), and we consider a maximum depth ($maxDepth$) of 3, meaning that at most three edges can be included in a translation path. The set of intermediate languages ($L$) includes five programming languages: Python, Rust, JavaScript, C++, and Go. 

ToCT (see Algorithm~\ref{alg:ToCT}) starts by enqueueing and then dequeueing the source PL, yielding the current path starting from the source PL, i.e., [Python] and the current depth 0 in our running example. Since Python is not the target language, and the current depth is less than the maximum depth of 3, the algorithm continuously explores possible transitions either to an intermediate language (excluding Python, since a PL cannot be translated to itself) or directly to the target language to complete the translation path. This results in the following paths: [Python, Java], [Python, Rust], [Python, JavaScript], [Python, C++], and [Python, Go]. Each of these new paths, along with the incremented depth of 1, is enqueued into $Q$. 
%That means new paths are generated by appending each language from the union set of ${Java}$ and set $\{$Rust, JavaScript, C++, Go$\}$ to the current path. 

Continuing this process, the algorithm dequeues [Python, Java] (i.e., the direct translation path) and since it ends with the target PL, this path will be added to the final translation PL path output list. Next, the algorithm dequeues [Python, Rust] and explores further transitions, appending each language from the set $L$ to the current path, but excluding Rust to avoid translation between the same PLs. This results in new paths like [Python, Rust, Java], [Python, Rust, JavaScript], etc., which are then enqueued with a depth of 2. This process repeats for all potential paths within the specified maximum depth, ensuring all possible translation paths from Python to Java are explored and recorded. By the end of the algorithm, the list $paths$ will contain all feasible sequences of translations from Python to Java, considering all given intermediate languages and the maximum depth argument.

\begin{algorithm}[h]
\caption{ToCT path generation algorithm}\label{alg:ToCT}
\begin{algorithmic}[1]
\REQUIRE $L_s$: Source programming language, $L_t$: Target programming language, $maxDepth$: Maximum depth of the tree, $L=\{L_i\}$: A set of intermediate languages.
\ENSURE All paths from $L_s$ to $L_t$
\STATE Initialize an empty list $paths$
\STATE Initialize a queue $Q$
\STATE Enqueue $([L_s], 0)$ into $Q$
\WHILE{$Q$ is not empty}
    \STATE $(currentPath, currentDepth) \leftarrow$ Dequeue $Q$
    \STATE $currentLang \leftarrow$ last element of $currentPath$
    \IF{$currentLang = targetLang$}
        \STATE Append $currentPath$ to $paths$
    \ELSIF{$currentDepth < maxDepth$}
        \FOR{$lang \in  \{L_t\} \cup L $}
            \IF{$lang \neq currentLang$}
                \STATE $newPath \leftarrow currentPath + [lang]$
                \STATE Enqueue $(newPath, currentDepth + 1)$ into $Q$
            \ENDIF
        \ENDFOR
    \ENDIF
\ENDWHILE
\STATE \textbf{return} $paths$
\end{algorithmic}
\end{algorithm}

\vspace{-0.2cm}
\subsection{Stage 2: Sequential Verification of ToCT}

For a specific translation problem (source program), the second stage of the \intertrans approach (see Algorithm~\ref{alg:verify}) takes the ToCT-generated plan for the problem's source and target PL, i.e., the list $paths$ from Algorithm~\ref{alg:ToCT}, to (1) determine the order of the paths that will be verified (i.e., checked if they lead to a successful translation), (2) generate the translations using an $LLM$ and a prompt template $PromptT$, and (3) evaluate the translations to the target language using the given test suite $T$. To make \intertrans more efficient, an \textit{early-stopping mechanism} is applied (Lines 19-20): as soon as one path successfully translates the code into $L_t$, Algorithm~\ref{alg:verify} terminates.

\begin{algorithm}
\caption{Algorithm for executing ToCT-generated plans}~\label{alg:verify}
\begin{algorithmic}[1]
\REQUIRE $P_s$: An input source program, $paths$: A list of translation PL paths generated by ToCT, $LLM$: a LLM that can generate code into $\{L_t\} \cup L$, $PromptT$: A prompt template for the specific LLM, $T$: a test suite for evaluating the computational accuracy of the generated translation to target PL $L_t$.
\ENSURE Successful translation, if any, from $L_s$ to $L_t$ for $P_s$
\STATE Sort $paths$ by their length in ascending order
\FOR{path $p \in paths$}
    \FOR{edge $E_k \in p$} 
        \IF{$E_k$ is already processed}
            \STATE \textbf{continue with cached output}
        \ELSE
        \STATE Retrieve extracted source code from $E_{k-1}$
        \STATE Create a new prompt using $PromptT$
        \STATE Perform translation using $LLM$ and the prompt
        \STATE Extract source code from inference output
        \IF{Failed extracting source code}
            \STATE \textbf{break} continue with the next path $p$
        \ENDIF
        \STATE Save the extracted code for $E_k$ to cache
        \ENDIF
        \IF{Target language of $E_k = L_t$} 
            \STATE Verify this translation using the test suite $T$
            \IF{Test suite passes}
                \STATE \textbf{return} the translation found
            \ENDIF
        \ENDIF
    \ENDFOR
    \STATE \textbf{return} {the translation failed}
\ENDFOR
\end{algorithmic}
\end{algorithm}

Following the design of ToCT, it is common for multiple paths to share the same initial transitive translation edges. For instance, Path $p_1$: [Python, Rust, Java] and Path $p_2$: [Python, Rust, JavaScript] Java share the first translation (edge). To further improve the efficiency of \intertrans, we apply memoization within each path to ensure the same edge is not computed more than once (Lines 4-5). Note that this optimization requires deterministic output for the same input prompt, which is ensured via a fixed seed in our experiment. Only new translation edges after branching from a shared path are processed. In other words, if $p_1$ is verified first, then $p_2$ will reuse the resulting Rust program saved in the memory cache to continue its unique translation to JavaScript.

In Algorithm~\ref{alg:verify}, the input $paths$ are first sorted by length in descending order (Line 1), ensuring that the first explored path is always a direct translation from $L_s$ to $L_t$. In the best-case scenario, where the LLM generates a program in the target language that successfully passes the evaluation test suite, the algorithm completes after exploring only this direct path. If no direct translation is found, the sorting step following path generation ensures that the algorithm maximizes the number of paths explored relative to the total translations performed. For instance, for our running example, in Figure~\ref{fig:tree-example} the numbers along the edges indicate the sequence of steps performed following Algorithm~\ref{alg:verify} for a specific $P_s$. The direct translation, i.e., [Python, Java], will be verified first. If the transferred code generated following this path fails, then the path [Python, Rust, Java] will be verified, and so on, until the transferred code generated by path [Python, C++, Java] passes the test suite $T$, the algorithm stops and returns the successful translation.

For each edge in a translation path, we first generate translated code for the target language of the previous edge (Line 7) (which serves as the source program of the current edge). Next, we use the given $LLM$ to generate the translation output (Lines 8-9), then extract the source code from this output (Line 10). If the extraction is successful, we then verify if it can pass the test suite $T$. 

%% file: exp.tex
We evaluate the effectiveness of \intertrans by answering the following three research questions:
\begin{itemize}
    \item RQ1: How effective is \intertrans compared to direct translation and other baselines?
    \item RQ2: How could varying the $maxDepth$ affect the performance of \intertrans?
    \item RQ3: How could varying the selection of intermediate languages affect \intertrans?  
\end{itemize}

\subsection{Benchmark Dataset Collection and Pre-Processing}
Our experiment dataset consists of 4,926 translation problems across 30 source-target translation PL pairs involving six PLs - C++, Go, Java, JavaScript, Python, and Rust. When creating our experiment dataset, we considered three existing datasets. Below, we describe the creation of our experimental datasets from these sources.

\noindent \textbf{TransCoder:} The original TransCoder dataset~\cite{roziere2020unsupervised} was created by manually collecting coding problems and solutions written in C++, Java, and Python from GeeksforGeeks~\cite{geeks4geeks}. Recently, Yang et al.~\cite{yang2024exploring} discovered quality issues in this dataset and subsequently conducted a manual verification and curation of the dataset to ensure its correctness. In this study, we reused their cleaned version, containing a total of 2,826 translation problems and corresponding test suites. We employed the full version of this dataset for comparisons with SOTA learning-based approaches.

\noindent \textbf{HumanEval-X:} HumanEval-X~\cite{zheng2023codegeex} extends the python-only code generation evaluation dataset HumanEval~\cite{chen2021evaluating} with additional canonical solutions and test cases in six PLs: C++, Go, Java, JavaScript, Python, and Rust. We created translation pairs for all 164 tasks in HumanEval-X across the six languages, resulting in 4,920 translation problems. Due to computational constraints (particularly required by the ablation studies performed to understand the impact of varying variables on the performance of \intertrans), we randomly sampled 1,050 translation problems, stratified across the 30 source-target translation pairs, ensuring a 99.9\% confidence level.

\noindent \textbf{CodeNet:} CodeNet~\cite{puri2021codenet} contains programs written in 55 programming languages for learning and evaluating coding tasks and was adopted in a recent empirical study by Pan et al.~\cite{pan_understanding_2023} on LLM introduced translation bugs. Programming tasks in CodeNet are verified by matching the program outputs with the expected results. For our study, we selected tasks with at least three test cases to ensure adequate test suite coverage, resulting in 1,112 programming tasks. From these tasks, we generated 15,660 translation problems by concentrating on the six PLs featured in HumanEval-X, removing problems with a file size exceeding 1KB (as a proxy for token length, to prevent inputting into the prompt problems longer than the model's token limit) and ensuring that each translated code snippet could be assessed using three test cases. We created a subset of 1,050 pairs from this dataset using stratified random sampling, ensuring a 99.9\% confidence level.

\subsection{Selected Large Language Models}
InterTrans relies on an LLM that understands multiple PLs. Almost all recent code LLMs possess this multilingual capability. We have chosen the following three instruct-tuned LLMs over their base models, as instruct-tuned models are fine-tuned to follow prompted instructions more effectively.

\noindent \textbf{Magicoder~\cite{wei_magicoder_2023}}: An open-source collection of LLMs trained on 75K synthetic instruction-response pairs and includes multiple model variants with different base models. All Magicoder models have around 7B parameters. We use the Magicoder-S-DS variant~\cite{magicoder-s-ds}. 

\noindent \textbf{StarCoder2~\cite{lozhkov2024starcoder}}: An open-source collection of LLMs offered by the BigCode project~\cite{BigCode}. StarCoder2 has instruction-tuned versions ranging from 1B to 34B parameters. We use the StarCoder2-15B variant~\cite{StarCoder2-15B}.

\noindent \textbf{CodeLlama~\cite{roziere2023code}}: An open-source collection of LLMs offered by Meta based on Llama 2, specialized in code generation, with 7B, 13B, and 34B parameters. We use the CodeLlama-13B variant~\cite{CodeLlama}.

We chose these models because of their proven effectiveness in code generation tasks and their open-source nature, which promotes accessibility and collaborative development. Additionally, we prioritized models compatible with efficient inference frameworks, i.e., vLLM~\cite{kwon_efficient_2023}, while also ensuring they work well with platforms such as the HuggingFace Text Generation Interface~\cite{wolf_huggingfaces_2020}. This ensures that our selected models are not only high-performing but also practically feasible for widespread use in both research and industry settings.

\subsection{Compared Approaches}\label{sec:baseline}
\noindent \textbf{Direct translation (CA@1 and CA@10):} We compare \intertrans with direct translation by evaluating performance with a single attempt (CA@1) and multiple attempts (CA@10). For CA@10, a single prompt is used to generate ten translation candidates. The translation is considered successful if any of these ten attempts result in a correct translation. Comparing with CA@1 reveals the additional opportunities \intertrans discovers via ToCT. Since \intertrans utilizes multiple translation paths, it inherently makes more than one attempt, making a comparison with CA@1 alone insufficient. Hence, to find a fair number of attempts (k) for direct translation, we analyzed how many attempts \intertrans required to achieve a successful translation across the experiments. On average, 3.9 attempts were needed, with 75\% of cases successful within two attempts and less than 0.1\% requiring between 59 and 83 attempts. Therefore, we chose CA@10 as a stronger baseline, allowing ten attempts with a high temperature setting to generate diverse variants and increase the chances of passing the test suite. The distribution of the number of attempts made by \intertrans in our experiments is presented in the supplementary material.
    
\noindent\textbf{Non-LLM SOTA approaches:} TransCoder~\cite{roziere2020unsupervised} is an unsupervised model pre-trained with cross-lingual language modeling, denoising auto-encoding, and back-translation, leveraging a vast amount of monolingual samples. TransCoder-IR~\cite{szafraniec_code_2023}, an incremental improvement, introduces the idea of using a low-level compiler Intermediate Representation (IR) to enhance translation performance. In addition to TransCoder's pretraining tasks, TransCoder-IR includes translation language modeling, translation auto-encoding, and IR generation. TransCoder-ST~\cite{roziere2021leveraging} is another enhanced version of TransCoder that uses automatically generated test cases to filter invalid translations, improving performance. These models are trained on only a few PLs, i.e., Python, C++, and Java.

\noindent \textbf{GPT-3.5 and its enhanced version:} GPT-3.5 is a powerful closed LLM provided by OpenAI that is capable of code generation. We consider the gpt-3.5-turbo-0613 version. UniTrans with GPT-3.5 is an enhanced version designed for code translation, proposed by Yang et al.~\cite{yang2024exploring}. UniTrans generates test cases to aid LLMs in repairing errors by integrating test execution error messages into prompts. Despite UniTrans with GPT-3.5 requiring additional program repair and extra test cases, we include it as a baseline since it represents the state-of-the-art performance on the TransCoder dataset.

\subsection{Evaluation Metric} 
Similar to recent studies on LLM-based code translation~\cite{pan_understanding_2023,yang2024exploring}, we adopt execution-based evaluation metrics, i.e., Computational Accuracy (CA)~\cite{roziere2020unsupervised}. CA assesses whether a transformed target program produces the same outputs as the source function when given identical inputs. CA on a benchmark is the ratio of translation problems that have correctly translated to the target language. We choose CA over text-based metrics like BLEU score because LLMs can produce valid translations that differ from human-written references; text-based metrics might be misleading when evaluating translated code against the reference, i.e., they can yield high scores despite the two code versions being functionally distinct~\cite{zhou_codebertscore_2023}. In our study, we aim to explore the effectiveness of intermediate PL translations in generating functionally equivalent programs instead of merely focusing on textual similarity.

\subsection{Implementation} 
Our scalable reference implementation of the \intertrans algorithms is written in Go and implemented as a client (Python) and server (engine written in Go) architecture that communicates over gRPC~\cite{grpc}. The \intertrans engine utilizes vLLM \cite{kwon_efficient_2023} as the inference engine, given its performance and dynamic batching capabilities. It queries vLLM endpoints using round-robin to achieve data parallelism during inference and distribute the computational load evenly. The computational infrastructure used for our experiments consists of 6x NVIDIA RTX A6000 GPUs on an AMD EPYC Server with 128 CPU cores.

To ensure deterministic inference results from vLLM across all experiments involving InterTrans, we randomly generated a fixed random seed for inference. We set the decoder parameters top-p to 0.95, top-k to 10, and the temperature to 0.7. When evaluating the baseline performance of direct translation with CA@1 and CA@10, we do not fix the seed to ensure we generate diverse candidates. The selection of top-p, top-k, and temperature aligns with recent studies on code LLMs~\cite{dilhara2024unprecedented}.

During our experiments, even after identifying a successful translation, we still continue to explore and verify all potential translation paths. While one would not do this in practice when using \intertrans, it was essential in our empirical study to collect comprehensive data on all translation paths needed for addressing our research questions, particularly RQ3 (impact of removing intermediate PLs). However, this does not impact the reported CA results for \intertrans. 

%% file: results.tex
\subsection{RQ1: Effectiveness of \intertrans}
\label{section:rq1}

\input{tables/rq1-table}
\input{tables/rq1-table-baseline}

\noindent \textbf{Approach:} In \intertrans, the $maxDepth$ is set to 4, allowing for a maximum of four translations (edges) in a translation PL path. This parameter enables us to explore various translation paths (with 85 maximum attempts). The six PLs of the CodeNet and HumanEval-X benchmarks, i.e., Python, C++, JavaScript, Java, Rust, and Go, serve as intermediate languages. While the TransCoder dataset includes only Python, C++, and Java, additional languages like Rust, JavaScript, and Go can be used as intermediates. This flexibility is possible because \intertrans does not verify the correctness of intermediate translations unless they result in a program written in the target language.
 
\noindent \textbf{Results:} Table \ref{tab:ca-intertrans} presents the comparison of \intertrans with direct translation (CA@1 and CA@10) across the three datasets, for the three base LLMs. We calculated both absolute and relative differences with CA@10, as the latter serves as a stronger direct translation baseline. Table \ref{table:rq1baseline} displays the comparison of \intertrans (with StarCoder2) against non-LLM SOTA approaches, GPT-3.5 and its enhanced version on the TransCoder dataset.

As shown in Table~\ref{tab:ca-intertrans}, \intertrans consistently surpasses direct translation (CA@1 and CA@10) across all three datasets and all studied LLMs. It achieves an absolute improvement of 18.3\% to 43.3\% compared to direct CA@10. Specifically, on CodeNet, \intertrans shows an average absolute improvement of 26.2\% for Code Llama, 38.3\% for Magicoder, and 43.3\% for StarCoder2 when compared to Direct (CA@10), the largest among three datasets for all three models. Overall, \intertrans with Magicoder performs the best, with the highest CA on both CodeNet and HumanEval-X (see grey filled cells in Table~\ref{tab:ca-intertrans}). On TransCoder, \intertrans with StarCoder2 performs the best with a CA of 93.8\%, slightly higher than \intertrans with Magicoder (90.8\%).

When comparing \intertrans with StarCoder2 (the best variant of \intertrans on TransCoder), to other state-of-the-art approaches on TransCoder, our approach outperforms all others across all six source-target PL pairs (see Table~\ref{table:rq1baseline}). The second best performance is achieved by UniTrans with GPT-3.5.
All the LLM-based approaches considered in Table \ref{table:rq1baseline} perform consistently better than the TransCoder models, further showcasing the promising potential of LLMs in automated code translation.

\subsection{RQ2: Impact of Varying $maxDepth$}

\begin{figure}[h]
    \centering
    \includegraphics[width=0.88\linewidth]{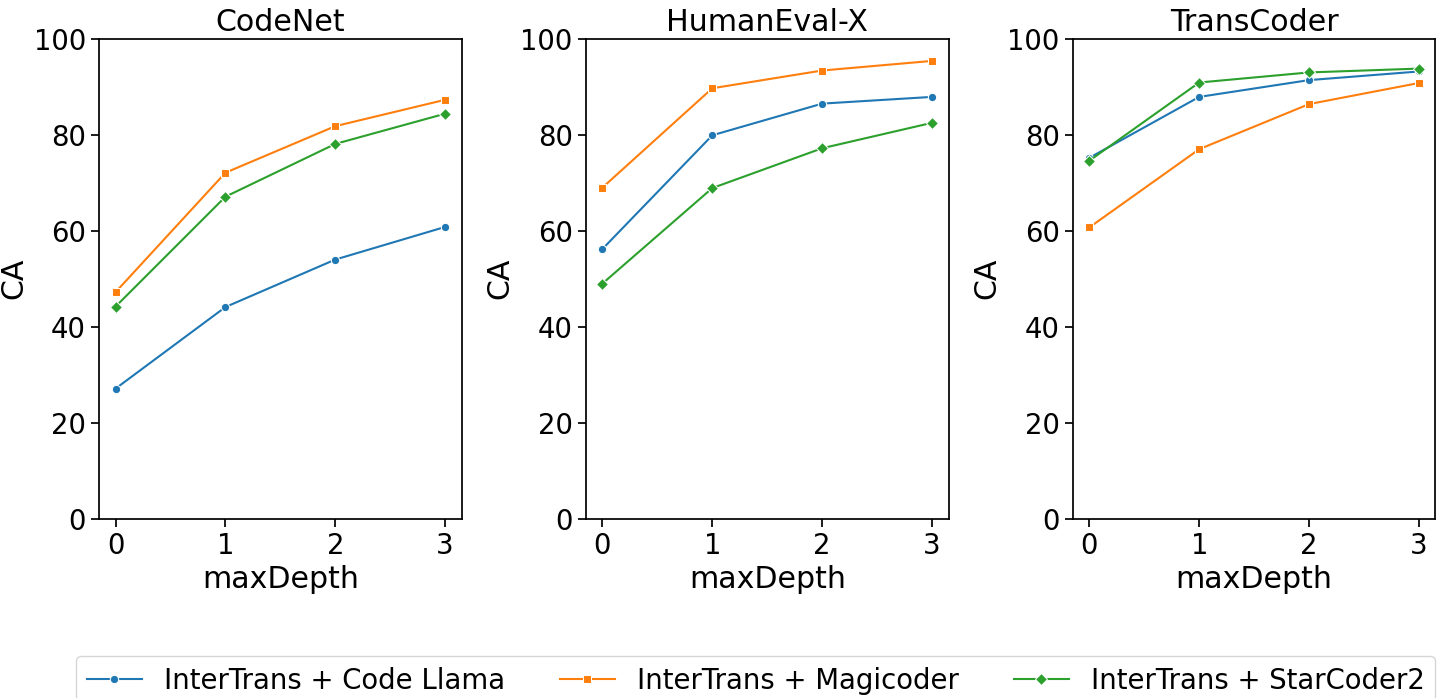} 
    \caption{Performance of \intertrans with varying $maxDepth$ on three datasets.}
    \label{fig:impactdepth}
\end{figure}

\noindent \textbf{Approach:} \intertrans utilizes two hyper-parameters, one of which is $maxDepth$. This parameter controls the depth of the translation tree generated by Algorithm~\ref{alg:ToCT}. In this research question, we investigate how this parameter affects the performance of \intertrans. Specifically, we vary $maxDepth$ from 1 (direct translation) to 4. We conducted pairwise comparisons across different depths (1 vs. 2, 1 vs. 3, 1 vs. 4, 2 vs. 3, 2 vs. 4, and 3 vs. 4) to evaluate the significance of the performance changes (i.e., the number of successful and unsuccessful translations) using the Chi-Square statistical test. To account for multiple comparisons across levels within the same model and dataset, we apply the Bonferroni correction to an alpha level of 0.05. The results of our experiments with varying values for $maxDepth$ are shown in Figure \ref{fig:impactdepth}. 

\noindent \textbf{Results:} We can observe that as the $maxDepth$ increases, the performance of \intertrans consistently improves, although the rate of improvement slows down towards longer paths. For instance, on HumanEval-X, increasing the $maxDepth$ from 1 to 2 results in an absolute improvement of 23.7\% for Code Llama, from 2 to 3 results in an improvement of 6.6\%, and from 3 to 4, the improvement is 3.2\%. Similar patterns are observed across all nine combinations of models and datasets.

Regarding the statistical tests performed, we find that for all datasets and models, there is a statistically significant improvement in terms of CA as the depth increases. Exceptions to this trend are noted for Code Llama and StarCoder 2 in the TransCoder dataset, where there is no significant increase in the CA metric when increasing the depth from 3 to 4, and for Code Llama and Magicoder in the HumanEval-X dataset with the same depth change. In other words, out of 54 comparisons (6 depth changes $\times$ 9) conducted, only 4 cases of increasing the depth do not lead to a statistically significant improvement in performance, all involving an increase from depth 3 to 4.

We observe small differences (an average of 2.8\%) when comparing the Direct (CA@1) reported in Table~\ref{tab:ca-intertrans} with the results of this experiment using $maxDepth$ set to 1. These differences may be attributed to the fact that for Direct CA@1 and CA@10, we did not fix the random seed. Consequently, a different seed was used.

\begin{figure*}[ht]
    \centering
    \includegraphics[width=\linewidth]{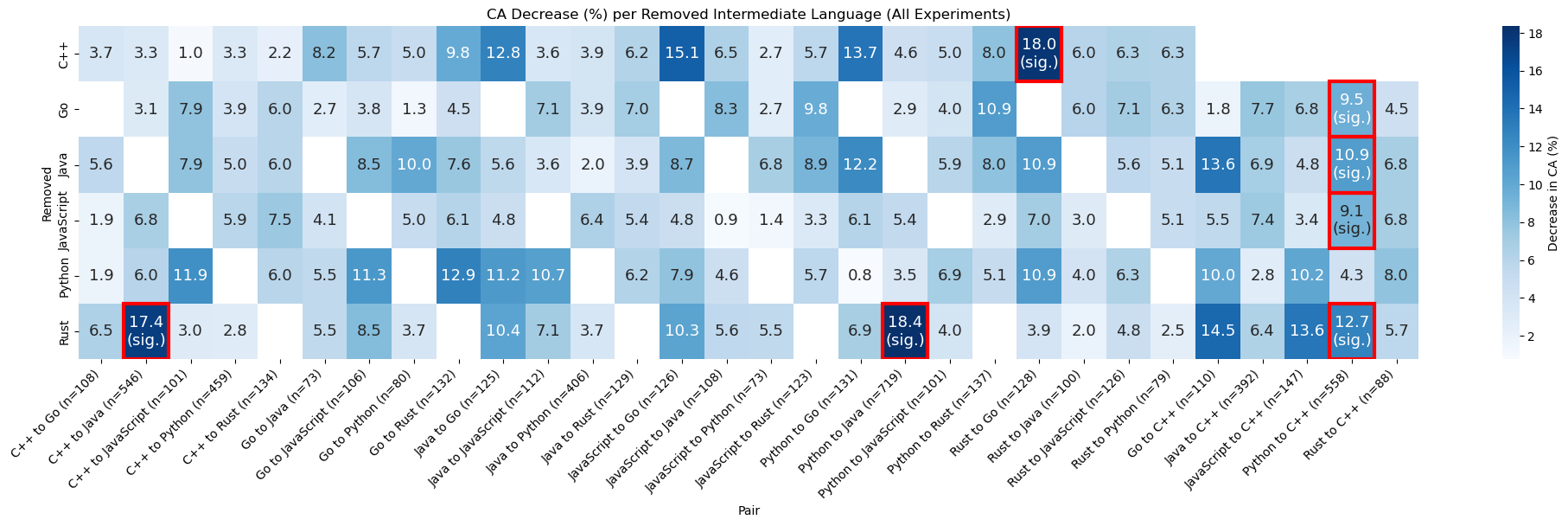} % Replace with your image file name
\caption{HeatMap showing the %relative 
mean absolute decrease in CA (\%) when removing a programming language from the intermediates used in our approach, compared to not removing any PL (across all datasets and models). Framed cells annotated with ``(sig.)'' indicate statistically significant results. The ``n'' value in the x-axis labels indicates the sample size for each translation pair. For each translation pair, one cell is empty because (by definition) the target PL can not be removed. }
    \label{fig:heatmap-languages}
\end{figure*}

\subsection{RQ3: Impact of Varying the Intermediate Programming Languages}\label{sec:rq3}
\noindent \textbf{Approach:} Besides $maxDepth$, the other hyper-parameter of \intertrans is the set of intermediate PLs considered, which determines the width of the translation tree created by ToCT. In this RQ, we investigate the impact of reducing the set and specific types of intermediate PLs by addressing the following two sub-RQs:

\begin{itemize}
\item \textbf{RQ3.1:} How does the number of available intermediate PLs influence the performance of \intertrans?
\item \textbf{RQ3.2:} How does the removal of a specific intermediate PL affect the performance of \intertrans?
\end{itemize}

To address the above two sub-RQs, we first conducted an ablation study across all possible combinations of intermediate PLs from the experiments conducted in RQ1 and RQ2, using a $maxDepth$ of 4 with six PLs. Each ablation involves the removal of all translation paths that contain a subset of the set of intermediate PLs. In particular, for each translation, we computed all 31 possible combinations of removing 1 to 5 PLs from the intermediates (i.e. all combinations of intermediate PLs, except those that include the target language). We then removed the edges that involve each individual set and measured whether the translation remained successful (i.e., at least one translation path leads to a correct translation). This ablation was performed for each sample of the nine experiments (3 datasets and 3 LLMs), and we recorded which removed sets caused the translation to be unsuccessful. For this analysis, we leveraged the data we generated during our evaluation described in Section~\ref{section:rq1}, where we recorded the execution result of all translation paths in the translation trees.

\begin{figure}[h]
    \centering
    \includegraphics[width=0.9\linewidth]{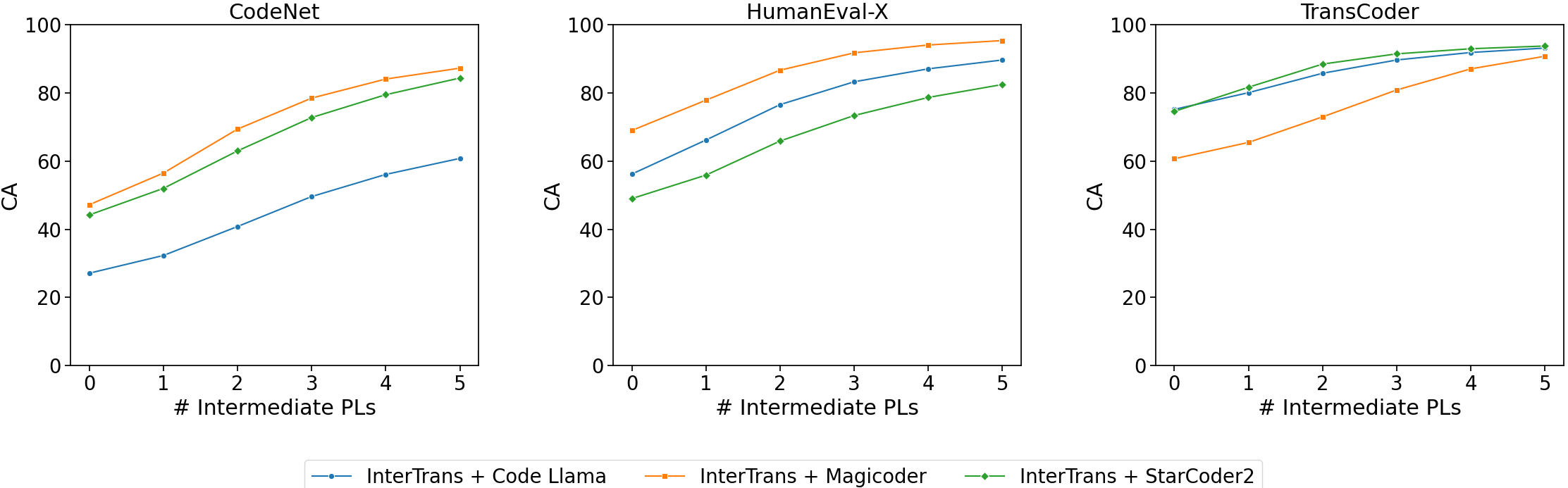} 
    \caption{Average performance of \intertrans with varying number of intermediate PLs on three datasets.}
    \label{fig:impactintermediatenum}
\end{figure}

To answer RQ3.1 in specific, we aggregated the results from the 458,118 translations (4,926 tasks from 3 datasets, each with 31 removal combinations using 3 different models) based on the number of intermediate PLs removed, i.e., the cardinality of the set of removed PLs. This analysis helps us understand the overall impact of the number of intermediate languages on translation success rate. Figure~\ref{fig:impactintermediatenum} shows the performance of \intertrans with 0 (direct translation) to 5 intermediate PLs on three datasets with three base models.

Additionally, in RQ3.2, to investigate whether specific languages are more impactful as intermediates, we analyzed the results from the translations of RQ3.1 that are associated with the removal of a single intermediate PL. We then calculated the mean absolute decrease in translation success for each of the 30 PL pairs in our experiments, caused by the removal of each specific PL. The heatmap in Figure~\ref{fig:heatmap-languages} shows the mean absolute decrease in CA when a PL is removed from each of the 30 translation pairs. Darker cells indicate a greater loss in CA, highlighting which PLs are more critical for maintaining high translation accuracy. This heatmap also shows the results of a statistical significance test (Chi-squared Goodness of Fit) we conducted by comparing the the number of successful and unsuccessful translations before (control group) and after (experimental group) the removal of a specific PL ($alpha = 0.05$, Bonferroni-corrected). In the heatmap, we highlight the cells associated with statistically significant differences.

\noindent \textbf{Results of RQ3.1:} We can observe in Figure~\ref{fig:impactintermediatenum} that the inclusion of more intermediate PLs consistently improves the translation accuracy of \intertrans. For instance, for Magicoder on CodeNet, increasing from zero to one intermediate PL results in a significant improvement of 9.3\% in CA (from 47.2\% to 56.5\%). Similarly, adding a second intermediate PL increases the CA metric by 12.9\%, and a third intermediate PL results in a 9.2\% increase. However, beyond this point, the incremental gains begin to diminish. Adding a fourth intermediate PL yields a 5.6\%, while the addition of a fifth intermediate PL results in a relatively smaller increase of 3.2\%. This trend suggests that while the inclusion of intermediate PLs is beneficial for improving translation accuracy, the marginal returns decrease as more intermediate PLs are added. The most substantial gains are observed when moving from zero to three intermediates, after which the improvements become more modest. 

\noindent \textbf{Results of RQ3.2:} 
Figure~\ref{fig:heatmap-languages} demonstrates that the importance of intermediate PLs varies across different translation pairs. For instance, when translating a program written in C++ to Java (second column of the heatmap), removing Rust as an intermediate PL resulted in a 17.4\% decrease in successful translations. In contrast, removing any other PL only led to a decrease ranging from 3.1\% to 6.8\%. This emphasizes the critical role of certain intermediate PLs in achieving accurate translations, yet we could not find any consistent trend across translation pairs.

To better understand the results of Figure~\ref{fig:heatmap-languages}, we conducted a case study in which we manually examined three translations associated with an absolute decrease in CA higher than 15\%. These cases were selected because their absolute decrease in CA were not only substantial but also statistically significant. Due to space limit, we only present the example translation for the first case, and include the details of three cases in the supplementary material.

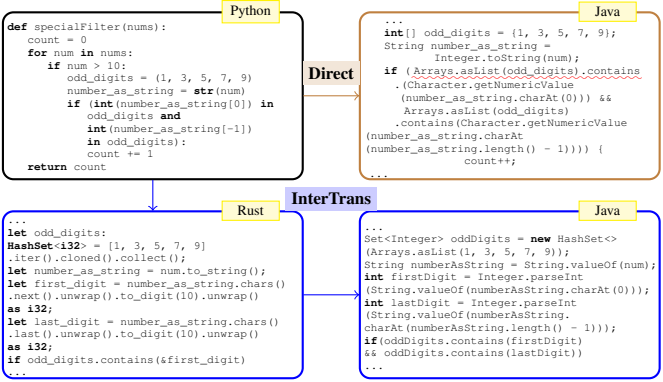
\begin{figure}[t]
\begin{center}
\begin{small}
\input{img/example-python-rust-java}
\caption{Example of translation from Python to Java via Rust.} \label{fig:example-python-rust-java}
\end{small}
\end{center}
\end{figure}

\noindent \textbf{Translation from Python to Java via Rust:} We observed that direct translation attempted to identify Java APIs and operations that are functionally equivalent to the Python ones, but which may not exist. Moreover, it struggles to handle type requirements in Java and thus often leads to the wrong use of API. Translating from Rust provided a pathway for translating these operations more accurately into Java. Figure~\ref{fig:example-python-rust-java} illustrates an example. In the source Python code, the expression "\texttt{if int(...) in []}" checks whether an integer is present in a list of integers. The direct translation uses the \texttt{Arrays.asList().contains()} API as an equivalent, but \texttt{Arrays.asList()} only accepts reference types, not primitive types. Consequently, passing an \texttt{int[]} to \texttt{Arrays.asList()} results in a \texttt{List<int[]>}, a list of arrays, failing to check for individual integers. In contrast, via the intermediate translation, the Rust code employs a \texttt{HashSet} for \texttt{odd\_digits}, which translates correctly to \texttt{Set<Integer>} in Java. This allows for accurate use of the \texttt{contains} method to check for individual elements.
    
\noindent \textbf{Translation from C++ to Java via Rust:} We found that direct translation frequently copies code from C++ to Java (since the two languages indeed have similar syntax), but some operations permitted in C++ are not permitted in Java, e.g., using square brackets to access vectors/lists and strings, implicit type conversion to lower precision types, etc. In \intertrans, C++ code is first translated into Rust with a more restrictive and distinct syntax, so that the LLM is aware of the syntax differences and avoids directly coping code.

\noindent \textbf{Translation from Rust to Go via C++:} Since Rust and Go both support type inference during variable declaration (i.e., developers can declare a variable without specifying the type if it can be inferred from the initialization expression), we observed that direct translation may misunderstand the types of local variables and try to apply invalid APIs on them.
For instance, direct translation attempts to translate Rust's \texttt{i32::abs} to Go's \texttt{Math.Abs}, without noticing that the return value (\texttt{w}) has changed from integer to float; this causes a type error three lines later at \texttt{w \% 10}. 
In \intertrans, the intermediate translation to C++ explicitly annotates the local variable \texttt{w} as int, and thus when the C++ is translated to Go, the LLM knows to wrap \texttt{w} with \texttt{int()} type conversion before performing the remainder operation.

%% file: tables/rq1-table.tex
\begin{table*}[h]
\caption{Performance of InterTrans compared with Direct Translation. Abs Diff and Rel Diff mean the absolute difference and relative difference compared to Direct (CA@10). The source language column includes all PLs of a dataset. The set of target languages for a given source language includes all PLs of a dataset, except the source language.}
\label{tab:ca-intertrans}
\resizebox{\textwidth}{!}{%
\begin{tabular}{@{}llrrrrrrrrrrrrrrrr@{}}
\toprule
\multirow[t]{3}{*}{\textbf{Dataset}} & \multirow[t]{3}{*}{\begin{tabular}[t]{@{}l@{}}\textbf{Source}\\ \textbf{language}\end{tabular}} & \multirow[t]{3}{*}{\begin{tabular}[t]{@{}l@{}}\textbf{Total}\\ \textbf{samples}\end{tabular}} & \multicolumn{15}{c}{\textbf{CA@K (percentage)}}                                                                                                                                                                                \\ \cmidrule(l){4-18} 
                         &                                                                            &                                                                          & \multicolumn{5}{l}{\textbf{Code Llama}}                                        & \multicolumn{5}{l}{\textbf{Magicoder}}                                         & \multicolumn{5}{l}{\textbf{StarCoder2}}                                         \\ \cmidrule(l){4-8} \cmidrule(l){9-13} \cmidrule(l){14-18}
                         &                                                                            &                                                                          & \begin{tabular}[t]{@{}l@{}}\textbf{Direct} \\ \textbf{(CA@1)}\end{tabular} & \begin{tabular}[t]{@{}l@{}}\textbf{Direct} \\ \textbf{(CA@10)}\end{tabular} & \begin{tabular}[t]{@{}l@{}}\textbf{\textsc{Inter}} \\ \textbf{\textsc{Trans}}\end{tabular} & \begin{tabular}[t]{@{}l@{}}\textbf{Abs.} \\ \textbf{diff.}\end{tabular} & \begin{tabular}[t]{@{}l@{}}\textbf{Rel.} \\ \textbf{diff.}\end{tabular} & \begin{tabular}[t]{@{}l@{}}\textbf{Direct} \\ \textbf{(CA@1)}\end{tabular} & \begin{tabular}[t]{@{}l@{}}\textbf{Direct} \\ \textbf{(CA@10)}\end{tabular} & \begin{tabular}[t]{@{}l@{}}\textbf{\textsc{Inter}} \\ \textbf{\textsc{Trans}}\end{tabular} & \begin{tabular}[t]{@{}l@{}}\textbf{Abs.} \\ \textbf{diff.}\end{tabular} & \begin{tabular}[t]{@{}l@{}}\textbf{Rel.} \\ \textbf{diff.}\end{tabular} & \begin{tabular}[t]{@{}l@{}}\textbf{Direct} \\ \textbf{(CA@1)}\end{tabular} & \begin{tabular}[t]{@{}l@{}}\textbf{Direct} \\ \textbf{(CA@10)}\end{tabular} & \begin{tabular}[t]{@{}l@{}}\textbf{\textsc{Inter}} \\ \textbf{\textsc{Trans}}\end{tabular} & \begin{tabular}[t]{@{}l@{}}\textbf{Abs.} \\ \textbf{diff.}\end{tabular} & \begin{tabular}[t]{@{}l@{}}\textbf{Rel.} \\ \textbf{diff.}\end{tabular} \\ \midrule
\multirow[t]{6}{*}{CodeNet} & C++                         & 175                     & 32.0                                           & 42.9                             & 61.1                & 18.3              & 42.7              & 50.3                            & 50.9                                               & 88.0                & 37.1                       & 73.0              & 29.1                            & 40.0                                               & 81.7                & 41.7                       & 104.3             \\
                                & Go                  & 175                     & 30.3                                           & 34.3                             & 61.1                & 26.9              & 78.3              & 50.9                            & 53.1                                               & 85.7                & 32.6                       & 61.3              & 45.7                            & 50.3                                               & 85.1                & 34.9                       & 69.3              \\
                                & Java           & 175                     & 25.7                                           & 38.9                             & 55.4                & 16.6              & 42.6              & 45.1                            & 45.7                                               & 85.1                & 39.4                       & 86.2              & 36.6                            & 41.1                                               & 85.7                & 44.6                       & 108.3             \\
                                
                                & JavaScript    & 175                     & 22.3                                           & 33.7                             & 64.6                & 30.9              & 91.5              & 50.9                            & 50.9                                               & 87.4                & 36.6                       & 71.9              & 24.0                            & 25.7                                               & 82.9                & 57.1                       & 222.2             \\
                                & Python     & 175                     & 14.3                                           & 19.4                             & 57.1                & 37.7              & 194.1             & 41.1                            & 42.3                                               & 91.4                & 49.1                       & 116.2             & 38.3                            & 44.0                                               & 87.4                & 43.4                       & 98.7              \\
                                & Rust       & 175                     & 29.7                                           & 38.3                             & 65.1                & 26.9              & 70.1              & 50.9                            & 51.4                                               & 86.3                & 34.9                       & 67.8              & 36.0                            & 45.1                                               & 83.4                & 38.3                       & 84.8              \\ \midrule
\multicolumn{2}{l}{\textbf{Total/Average}}          & 1,050           & 25.7                                           & 34.6                             & \textbf{60.8}                & 26.2              & 75.8              & 48.2                            & 49.0                                               & \cellcolor{black!15}\textbf{87.3}                & 38.3                       & 78.1              & 35.0                            & 41.0                                               & \textbf{84.4}                & 43.3                       & 105.6 \\ \midrule
            \multirow[t]{6}{*}{HumanEval-X}        & C++                        & 175                     & 70.3                                           & 78.9                             & 91.4                & 12.6              & 15.9              & 73.1                            & 74.3                                               & 97.7                & 23.4                       & 31.5              & 61.1                            & 66.9                                               & 86.3                & 19.4                       & 29.1              \\
                                & Go   & 175                     & 64.0                                           & 71.4                             & 90.3                & 18.9              & 26.4              & 62.9                            & 64.0                                               & 98.3                & 34.3                       & 53.6              & 52.0                            & 55.4                                               & 83.4                & 28.0                       & 50.5              \\
                                & Java     & 175                     & 58.3                                           & 68.0                             & 87.4                & 19.4              & 28.6              & 65.7                            & 67.4                                               & 93.1                & 25.7                       & 38.1              & 46.9                            & 48.6                                               & 86.3                & 37.7                       & 77.6              \\
                                & JavaScript   & 175                     & 57.1                                           & 73.1                             & 93.1                & 20.0              & 27.3              & 60.6                            & 60.6                                               & 96.0                & 35.4                       & 58.5              & 44.0                            & 44.0                                               & 80.6                & 36.6                       & 83.1              \\
                                & Python   & 175                     & 53.7                                           & 64.6                             & 82.3                & 17.7              & 27.4              & 61.7                            & 62.9                                               & 89.7                & 26.9                       & 42.7              & 36.6                            & 36.6                                               & 77.1                & 40.6                       & 110.9             \\
                                & Rust    & 175                     & 59.4                                           & 72.0                             & 93.7                & 21.7              & 30.2              & 71.4                            & 72.0                                               & 97.7                & 25.7                       & 35.7              & 52.6                            & 54.3                                               & 81.1                & 26.9                       & 49.5              \\ 
                                \midrule
\multicolumn{2}{l}{\textbf{Total/Average}}                & 1,050           & 60.5                                           & 71.3                             & \textbf{89.7}                & 18.4              & 25.8              & 65.9                            & 66.9                                               & \cellcolor{black!15}\textbf{95.4}                & 28.6                       & 42.7              & 48.9                            & 51.0                                               & \textbf{82.5}                & 31.5                       & 61.9              \\

\midrule

\multirow[t]{6}{*}{TransCoder}    & C++                  & 946                     & 73.9                                           & 75.9                             & 93.2                & 17.3              & 22.8              & 67.9                            & 67.9                                               & 92.7                & 24.8                       & 36.6              & 63.5                            & 65.2                                               & 93.8                & 28.5                       & 43.8              \\
                                & Java                      & 931                     & 77.7                                           & 79.5                             & 94.8                & 15.4              & 19.3              & 77.4                            & 77.4                                               & 91.9                & 14.5                       & 18.7              & 79.3                            & 79.9                                               & 95.1                & 15.1                       & 19.0              \\
                                & Python                       & 949                     & 67.3                                           & 69.3                             & 91.6                & 22.2              & 32.1              & 33.5                            & 33.5                                               & 87.8                & 54.3                       & 161.9             & 73.9                            & 74.6                                               & 92.7                & 18.1                       & 24.3              \\  \midrule
\multicolumn{2}{l}{\textbf{Total/Average}}          & 2,826 & 72.9                                           & 74.9                             & \textbf{93.2}                & 18.3              & 24.5              & 59.5                            & 59.5                                               & \textbf{90.8}                & 31.3                       & 52.6              & 72.2                            & 73.2                                               & \cellcolor{black!15}\textbf{93.8}                & 20.6                       & 28.2 \\

\bottomrule
\end{tabular}%
}
\end{table*}

%% file: tables/rq1-table-baseline.tex
\begin{table}[h]
    \centering
    \caption{CA performance of \intertrans and other baselines on TransCoder data set. We adopt the numbers of baseline performance from Yang et al.~\cite{yang2024exploring}. A ``--'' means there is no reported performance on the specific pair.}
    \label{table:rq1baseline}
     \resizebox{\linewidth}{!}{%
     \begin{tabular}{@{}lrrrrrrr@{}}
      \toprule
        \textbf{Models} &  \begin{tabular}[t]{@{}l@{}}\textbf{C++ to} \\ \textbf{Python}\end{tabular} & \begin{tabular}[t]{@{}l@{}}\textbf{Python} \\ \textbf{to C++}\end{tabular} & \begin{tabular}[t]{@{}l@{}}\textbf{Java} \\ \textbf{to C++}\end{tabular} & \begin{tabular}[t]{@{}l@{}}\textbf{C++ to} \\ \textbf{Java}\end{tabular} & \begin{tabular}[t]{@{}l@{}}\textbf{Java to} \\ \textbf{Python}\end{tabular} & \begin{tabular}[t]{@{}l@{}}\textbf{Python} \\ \textbf{to Java}\end{tabular} & \textbf{Avg.} \\ \midrule
        TransCoder & 36.6 & 30.4 & 27.8 & 49.8 & -- & -- & 36.2 \\
        TransCoder-IR  & -- & -- & 41.0 & 40.5 & -- & -- & 45.8 \\
        TransCoder-ST  & 46.3 & 47.8 & 49.7 & 64.7 & -- & -- & 52.2 \\
         GPT-3.5 &  87.1 & 89.5 & 92.9 & 82.2 & 89.2 & 74.9 & 86.0 \\
        UniTrans w/ GPT-3.5  & 88.8 & 94.2 & 94.9 & 85.5 & 91.2 & 81.3 & 87.9 \\ \midrule
        InterTrans w/ StarCoder2&  \textbf{93.3} & \textbf{94.4}  & \textbf{96.1}  & \textbf{94.2}  & \textbf{94.0}  & \textbf{91.1}  & \textbf{93.8} \\ \bottomrule
    \end{tabular}
    }
\end{table}

%% file: img/example-python-rust-java.tex
%% id_request = 807, model = magicoder, dataset = humanevalx
%% Python -> Java vs. Python -> Rust -> Java

\newsavebox\boxExamplePythonRustJavaSrc
\begin{lrbox}{\boxExamplePythonRustJavaSrc}
\begin{lstlisting}[language=java-pretty-no-number, language=python]
def specialFilter(nums):
    count = 0
    for num in nums:
        if num > 10:
            odd_digits = (1, 3, 5, 7, 9)
            number_as_string = str(num)
            if (int(number_as_string[0]) in 
                odd_digits and 
                int(number_as_string[-1]) 
                in odd_digits):
                count += 1
    return count 
\end{lstlisting}
\end{lrbox}

\newsavebox\boxExamplePythonRustJavaDirect
\begin{lrbox}{\boxExamplePythonRustJavaDirect}
\begin{lstlisting}[language=java-pretty-no-number, language=java]
    ...
    int[] odd_digits = {1, 3, 5, 7, 9};
    String number_as_string = 
              Integer.toString(num);
    if ((*@\!\!\coloruwave{red}{Arrays.asList(odd\_digits).contains}@*)
      .(Character.getNumericValue
       (number_as_string.charAt(0))) &&
        Arrays.asList(odd_digits)
      .contains(Character.getNumericValue
(number_as_string.charAt
(number_as_string.length() - 1)))) {
                    count++;
 ...
\end{lstlisting}
\end{lrbox}

\newsavebox\boxExamplePythonRustJavaInterTransRust
\begin{lrbox}{\boxExamplePythonRustJavaInterTransRust}
\begin{lstlisting}[language=java-pretty-no-number, language=rust]
...
let odd_digits: 
HashSet<i32> = [1, 3, 5, 7, 9]
.iter().cloned().collect();
let number_as_string = num.to_string();
let first_digit = number_as_string.chars()
.next().unwrap().to_digit(10).unwrap() 
as i32;
let last_digit = number_as_string.chars()
.last().unwrap().to_digit(10).unwrap() 
as i32;
if odd_digits.contains(&first_digit) 
...
\end{lstlisting}
\end{lrbox}

\newsavebox\boxExamplePyRustJava
\begin{lrbox}{\boxExamplePyRustJava}
\begin{lstlisting}[language=java-pretty-no-number, language=java]
...
Set<Integer> oddDigits = new HashSet<>
(Arrays.asList(1, 3, 5, 7, 9));
String numberAsString = String.valueOf(num);
int firstDigit = Integer.parseInt
(String.valueOf(numberAsString.charAt(0)));
int lastDigit = Integer.parseInt
(String.valueOf(numberAsString.
charAt(numberAsString.length() - 1)));
if(oddDigits.contains(firstDigit) 
&& oddDigits.contains(lastDigit)) 
...
\end{lstlisting}
\end{lrbox}

\DefMacro*{color-Direct}{brown}
\DefMacro*{color-InterTrans}{blue}

\begin{tikzpicture}[
  code box/.style={rectangle, thick, rounded corners, draw=black, minimum width=21em, minimum height=26ex},
  lang anno/.style={rectangle, fill=yellow!20, draw=yellow, minimum width=4em, minimum height=2ex},
  tool anno/.style={font=\bfseries, scale=1.2},
  scale=0.6,
  every node/.style={scale=0.6},
]

\coordinate (c-Src) at (0, 0);
\coordinate (c-Direct) at (25em, 0);
\coordinate (c-InterTransRust) at (0, -31ex);
\coordinate (c-InterTransJava) at (25em, -31ex);

\node (b-Src) [below = 0 of c-Src] [code box] {};
\node (b-Src-code) [right = 0 of b-Src.west] {\usebox{\boxExamplePythonRustJavaSrc}};
\node (b-Src-anno) [left = 1em of b-Src.north east] [lang anno] {Python};

\node (b-Direct) [below = 0 of c-Direct] [code box, draw=\UseMacro{color-Direct}] {};
\node (b-Direct-code) [right = 0 of b-Direct.west] {\usebox{\boxExamplePythonRustJavaDirect}};
\node (b-Direct-anno) [left = 1em of b-Direct.north east] [lang anno] {Java};

\draw[->, draw=\UseMacro{color-Direct}] (b-Src.east) -> (b-Direct.west);
\coordinate (c-Direct-mid) at ($(b-Src.east)!0.5!(b-Direct.west)$);
\node (b-Direct-tool-anno) [above = 1ex of c-Direct-mid] [tool anno, fill=\UseMacro{color-Direct}!20] {Direct};

\node (b-InterTransRust) [below = 0 of c-InterTransRust] [code box, draw=\UseMacro{color-InterTrans}] {};
\node (b-InterTransRust-code) [right = 0 of b-InterTransRust.west] {\usebox{\boxExamplePythonRustJavaInterTransRust}};
\node (b-InterTransRust-anno) [left = 1em of b-InterTransRust.north east] [lang anno] {Rust};

\node (b-InterTransJava) [below = 0 of c-InterTransJava] [code box, draw=\UseMacro{color-InterTrans}] {};
\node (b-InterTransJava-code) [right = 0 of b-InterTransJava.west] {\usebox{\boxExamplePyRustJava}};
\node (b-InterTransJava-anno) [left = 1em of b-InterTransJava.north east] [lang anno] {Java};

\draw[->, draw=\UseMacro{color-InterTrans}] (b-Src.south) -> (b-InterTransRust.north);
\draw[->, draw=\UseMacro{color-InterTrans}] (b-InterTransRust.east) -> (b-InterTransJava.west);
\coordinate (c-InterTrans-northmid) at ($(b-InterTransRust.north)!0.5!(b-InterTransJava.north)$);
\node (b-InterTrans-tool-anno) [above = 0 of c-InterTrans-northmid] [tool anno, fill=\UseMacro{color-InterTrans}!20] {InterTrans};
    
\end{tikzpicture}

%% file: discuss.tex
\begin{comment}

\begin{algorithm}
\caption{Verify Algorithm for Code Translation}
\begin{algorithmic}[1]
\REQUIRE $P = (T_1, T_2, \ldots, T_n)$ list of translation paths.
\ENSURE Successful translation from $A$ to $B$
\STATE Sort $P$ by path length in ascending order
\FOR{each path $p \in P$}
    \STATE Let $p = (E_1, E_2, \ldots, E_n)$
    \FOR{$k = 1$ to $n$}
        \IF{Edge is already processed}
            \STATE continue
        \ENDIF
        \STATE Retrieve extracted source code from previous edge
        \STATE Build prompt
        \STATE Perform translation for this translation edge
        \STATE Extract source code from inference output
        \STATE Verify this translation using the test suite
        \IF{Test suite passes}
            \IF{Target language of $E_i$ = $B$}
                \STATE \textbf{return} the translation found
            \ENDIF
        \ELSE
            \STATE \textbf{continue} {next path}
        \ENDIF
    \ENDFOR
\ENDFOR
\end{algorithmic}
\end{algorithm}
\end{comment}

\subsection{Implications}

\noindent \textbf{\intertrans vs. Other Tools:} We demonstrate that \intertrans is a better alternative to state-of-the-art approaches (see Table \ref{tab:ca-intertrans}). By leveraging easily accessible open-source code LLMs, such as Magicoder, \intertrans achieved an average CA ranging from 87.3\%-95.4\% across three datasets involving 30 different source-target PL pairs. Furthermore, \intertrans requires only readily available LLMs and a relatively limited depth and set of intermediate languages to perform effectively (see Figures \ref{fig:impactdepth} and \ref{fig:impactintermediatenum}). 

\noindent \textbf{Computational Cost and Efficiency of \intertrans}: Despite implementing various optimizations in \intertrans, such as caching inferences of shared edges, the system remains computationally expensive. This is largely due to the ToCT algorithm, which explores multiple translation paths, each potentially containing numerous translations. However, the significant translation performance gains suggest that \intertrans can save considerable time compared to alternatives, especially in contexts where human resources are costly. Additionally, by leveraging existing LLMs, we avoid the expense of training a specific code translation model. In practice, the actual cost of \intertrans is much lower than the theoretical maximum, as our experiments indicate an average of only 3.9 attempts per successful translation (see Section~\ref{sec:baseline}). Future research can improve its efficiency by parallelizing the currently sequential inference process and developing methods to predict the most likely successful path for specific translation problems instead of iteratively evaluating different paths.

\noindent \textbf{Multiple and Dynamic Intermediates vs. Unified IR}: Our study confirms that prior work was on the right track by utilizing intermediate representations. However, our approach innovates by employing multiple, dynamic intermediates tailored to each source-target language pair, utilizing existing PLs instead of a unified compiler-level representation like LLVM IR. Our findings suggest that a single, fixed intermediate language may not suffice, as the performance impact varies depending on the languages involved (see Figure~\ref{fig:heatmap-languages}). Even though each successive intermediate translation can potentially increase the risk of propagating translation errors to the next translation, in practice this risk turned out to be moderate, with substantial \emph{improvements} of translation quality. Through an initial analysis of three translation patterns, we uncovered several interesting insights (see Section~\ref{sec:rq3}). However, future work is needed to understand why certain paths are more effective in particular scenarios and to develop methods for recommending the optimal translation path for a given translation problem.

\subsection{Threats to Validity}

\noindent \textbf{Internal Validity:} We performed the translation only once for each translation problem, using a fixed random seed for study LLMs when reporting the performance of \intertrans. This design reduces the risk of selecting a favorable seed across all nine experiments. However, altering this seed could affect the reported performance. Nonetheless, this does not affect the comparison between \intertrans with direct translation (as shown in Table \ref{fig:impactdepth}, where depth=1 represents direct translation under identical conditions), or the empirical analysis of varying parameters, which are our main goals. 

Furthermore, we employed a single prompt template for each dataset; changing this template might also alter the reported performance across all models. However, this does not affect comparison results, as we used the same prompt for all models, including direct translation and \intertrans. To mitigate the effects of LLMs' sensitivity to prompt templates, we adhered to best practices from the literature. Future practitioners can explore potential improvements in prompt design. 

The temperature and top-p, top-k values were set consistently across all LLMs, following established literature. While these may not be the optimal parameters for a specific model, our primary objective is to demonstrate the improvement of \intertrans over direct translation, regardless of the LLMs used.

Another threat to internal validity arises from potential data leakage in LLMs, meaning there could be an overlap between the training data of the studied LLMs and the evaluation dataset used in this work. However, this issue would impact all baseline models, not just \intertrans, ensuring that the relative performance comparisons between models in our study remain valid. Additionally, unlike code generation, open-source code corpora typically do not contain paired code translation data (i.e., source and target code in a single file). We also carefully reviewed the documentation for Magicoder and StarCoder2 and they did not include code translation as a fine-tuning task.

\noindent \textbf{External Validity:} Potential threats to external validity may arise from the selection of target PLs, LLMs, evaluation datasets, and compared approaches. To mitigate these threats, we selected six popular PLs with varying levels of maturity, encompassing different programming paradigms. The source-target PL pairs we considered include all those concerned in recent work on LLM-based code generation by Pan et al.~\cite{pan_understanding_2023} and Yang et al.~\cite{yang2024exploring}. For dataset selection, our evaluation set is sourced from three well-known benchmarks. Two of these benchmarks were used in the previously mentioned studies, and the third allows for a fair comparison with non-LLM-based models, such as the TransCoder family and GPT-3.5. We selected three popular and recent open-source code LLMs as the base for \intertrans. These models are multilingual and have demonstrated strong performance on code generation tasks. In the future, additional LLMs can be seamlessly integrated, as \intertrans's implementation is LLM-agnostic, meaning all LLMs will be treated equally without requiring additional engineering steps.

\noindent \textbf{Construct Validity:} Similar to prior studies\cite{pan_understanding_2023,yang2024exploring}, we only consider execution-based evaluation metric, i.e., CA. While execution-based metrics align better with our goal to investigate the capability of LLMs in generating translated code that is functionally equal to the source program, its reliability will be impacted by the effectiveness of output control and the quality of test cases. To mitigate these threats, we applied output control following the best practices suggested by Macedo et al.~\cite{macedo2024exploring} and calculated the matching success rate (MSR) in extracting source code from inference output for all nine experiments. These values range from 97.7\% to 100\%, with an average MSR of 99.7\%. This indicates that the reported performance is unlikely to be significantly influenced by the varying capabilities of the LLMs in generating source code that can be automatically extracted. 

For the evaluation datasets, we used the complete, cleaned TransCoder dataset, allowing us to leverage the performance metrics reported by Yang et al.~\cite{yang2024exploring} for SOTA approaches, where each translation includes at least one test case. When sampling from the CodeNet dataset, we ensured that each translation problem has at least three tests. Additionally, each translation problem in HumanEval-X contains an average of 7.7 tests. 

%% file: related.tex
\subsection{Automated Code Translation}\label{sec:related_code_translation}
Automated code translation approaches generally fall into two categories: rule-based methods and data-driven learning-based methods. Rule-based automated code translation approaches~\cite{c2rust,cxgo,sharpen,j2cstranslator} utilize program analysis techniques and handcrafted rules to translate code between programming languages (PLs). A prominent example is C2Rust~\cite{c2rust}, which has gained significant attention with 3.8k stars on GitHub as of this writing. However, these tools often produce non-idiomatic translations and are expensive to develop~\cite{szafraniec_code_2023}. Learning-based approaches aim to address these limitations by leveraging large-scale data. These methods can be supervised, using parallel code translation pairs, or unsupervised, learning from open-source code. Techniques in this category have evolved significantly, starting with statistical learning techniques~\cite{nguyen2013lexical,nguyen2014migrating,karaivanov2014phrase}, progressing to neural network approaches~\cite{chen2018tree}, and more recently, to pre-trained model-based~\cite{lachaux2021dobf,roziere2021leveraging,szafraniec_code_2023,jiao_evaluation_nodate,pan_understanding_2023} and LLM-based methods~\cite{pan_understanding_2023,yang2024exploring}. Our proposed \intertrans is also a LLM-based code translation approach. It is unique among existing methods as it is the first study to explore the potential of leveraging intermediate PLs for code translation.

\subsection{The Multilingual Capacity of Code LLMs}
Code LLMs such as Codex~\cite{chen2021evaluating}, CodeGen~\cite{nijkamp2022codegen}, Code Llama~\cite{roziere2023code}, and StarCoder2~\cite{lozhkov2024starcoder}, have shown great potential in tasks like code understanding, summarization, and generation. These models, especially those with a large number of parameters (e.g., 7B, 13B, or larger), are pre-trained on extensive code databases using self-supervised autoregressive objectives. All these models are trained on multiple PLs to ensure their generalizability across various coding tasks, thereby possessing multilingual capabilities. For example, StarCoder2~\cite{lozhkov2024starcoder} models are trained on Stack v2, which is built on the Software Heritage’s vast source code archive spanning over 600 PLs~\cite{SoftwareHeritage2024} and other high-quality open data sources such as GitHub issues, pull requests, etc.

HumanEval~\cite{chen2021evaluating}, developed by OpenAI, is the first benchmark for assessing the code generation capabilities of LLMs. However, this dataset only contains Python coding problems and does not assess the multilingual capabilities of code LLMs. Recently, efforts have increased to empirically evaluate the multilingual capabilities of code LLMs~\cite{zheng2023codegeex,cassano2023multipl,peng2024humaneval,chai2024mceval}. For instance, Zheng et al.~\cite{zheng2023codegeex} evaluated their multilingual code model, CodeGeeX, pre-trained on code written in 23 PLs, on their developed new benchmark HumanEval-X. HumanEval-X extends HumanEval to include handwritten solutions and test cases for five additional PLs: Rust, Go, JavaScript, C++, and Java. Recent studies by Pan et al.~\cite{pan_understanding_2023} and Yang et al.~\cite{yang2024exploring} conducted comprehensive empirical evaluations of LLMs' capabilities in automated code translation, including programs written in multiple PLs and leveraging multilingual benchmarks like CodeNet~\cite{puri2021codenet} and TransCoder~\cite{roziere2020unsupervised}. Our experiments also include these two datasets. However, previous work has not leveraged the multilingual capability of code LLMs for the inference of code translation, focusing instead on evaluation.

%% file: conclude.tex
This work explores the potential of leveraging the multilingual capabilities of LLMs to enhance automated code translation through transitive intermediate translations. We propose \intertrans, a novel approach that utilizes a planning algorithm (ToCT) to generate candidate translation paths, which are then evaluated sequentially. Through extensive empirical studies on three benchmarks, our results demonstrate the promise of \intertrans with an \textbf{absolute improvement boosting of 18.3\% to 43.3\%} in Computation Accuracy (CA) over direct translation with ten attempts. With only a readily available open-source LLM, e.g., Magicoder, \intertrans achieved an average CA of 87.3\%-95.4\% on three benchmark datasets. \intertrans not only enhances translation accuracy, but also provides a new direction for future research in leveraging and interpreting multilingual LLMs for diverse coding tasks.